\documentclass{PoS}

\usepackage{epsfig}
\usepackage{latexsym}
\usepackage{graphicx}
\usepackage{bm}
\usepackage{longtable}

\def\lsim{\raise0.3ex\hbox{$<$\kern-0.75em\raise-1.1ex\hbox{$\sim$}}}
\def\gsim{\raise0.3ex\hbox{$>$\kern-0.75em\raise-1.1ex\hbox{$\sim$}}}
\def\simgt{\rlap{\lower 3.5 pt\hbox{$\mathchar \sim$}}\raise 1.0pt \hbox {$>$}}
\def\simlt{\rlap{\lower 3.5 pt\hbox{$\mathchar \sim$}}\raise 1.0pt \hbox {$<$}}

\title{Bound state of two-nucleon systems in quenched lattice QCD}

\ShortTitle{Bound state of two-nucleon systems in quenched lattice QCD}

\author{Takeshi Yamazaki
\ for PACS-CS Collaboration
\\ \\
Kobayashi-Maskawa Institute for the Origin
 of Particles and the Universe, Nagoya University, Naogya, Aichi 464-8602, 
Japan\\
        \email{yamazaki@kmi.nagoya-u.ac.jp}
}

\abstract{
We address the issue of bound state in the two-nucleon system in lattice QCD
with the quenched approximation at the lattice spacing of $a =0.128$ fm
using a heavy quark mass corresponding to $m_\pi = 0.8$ GeV.
To distinguish a bound state from an attractive scattering state,
we investigate the volume dependence of the energy difference 
between the ground state and the free two-nucleon state by changing the
spatial extent of the lattice from 3.1 fm to 12.3 fm.
A finite energy difference left in the infinite spatial volume limit
leads us to the conclusion that the measured ground states for 
not only spin triplet but also singlet channels
are bounded. Furthermore the existence of the bound state is confirmed by
investigating the properties of the energy for the first excited state
obtained by 2$\times$2 diagonalization method.
The scattering lengths for both channels are evaluated by
the finite volume formula derived by L\"uscher.
}

\FullConference{The XXIX International Symposium on Lattice Field Theory - Lattice 2011\\
July 10-16, 2011\\
Squaw Valley, Lake Tahoe, California}

\begin{document}

\section{Introduction}
\label{sec:introduction}

The strong interaction dynamically generates
a  hierarchical structure: three quarks are bound to form a nucleon with an energy of 1~GeV,  
and nucleons are in turn bound to form nuclei with a binding energy of 10~MeV or so per nucleon. 
This is a multi-scale physics that computational physics should explore, and lattice QCD is responsible 
for explaining the nature of nuclei based on first principles.

Recently, we have made a first attempt to directly construct 
the Helium-3 and Helium-4 nuclei 
in quenched QCD~\cite{Yamazaki:2009ua} at a rather heavy quark mass 
corresponding to $m_\pi=0.8$ GeV, and 
successfully confirmed the formation of Helium nuclei as a bound state.  
After our finding of the Helium nuclei, an evidence of the H di-baryon bound 
state in $N_f = 2 + 1$ and $N_f = 3$ QCD were reported by
NPLQCD~\cite{Beane:2010hg,Beane:2011xx} and HALQCD~\cite{Inoue:2010es} 
Collaborations, respectively.
The situation, however, is markedly different for deuteron, 
which is the simplest nucleus composed of 
two nucleons in the spin triplet channel, and yet 
evidence based on lattice QCD for bound state had never been reported
before our paper~\cite{Yamazaki:2011xx}. 
It is already quite some time ago that  a first analysis of the two-nucleon system was made 
in quenched QCD~\cite{Fukugita:1994na}.
Much more recently, studies were made with a partially-quenched 
mixed action~\cite{Beane:2006mx}
and $N_f = 2+1$ anisotropic Wilson action~\cite{Beane:2009py}.
Extraction of the potential between two nucleons has been investigated in
quenched QCD~\cite{Ishii:2006ec}. 
All these studies, however, tried to calculate the two-nucleon scattering
lengths assuming, based primarily on model considerations with nuclear potentials, 
 that the deuteron becomes unbound for 
the heavy quark mass, corresponding to $m_\pi \simgt \ 0.3$ GeV, 
employed in their simulations.

To check the validity of this assumption,
we need to investigate whether the bound state exists or not in the
heavy quark mass region, where studies  so far have been carried out, using the arsenal of 
methods solely within lattice QCD.  
If there is a bound state, the ground state energy never yield
the scattering length if substituted into the L\"uscher's finite volume 
formula~\cite{Luscher:1986pf}.
In such a  case, the scattering length should be obtained from 
the energy of the first excited state.

We carry out two types of calculations at a heavy quark mass 
corresponding to $m_\pi = 0.8$ GeV in quenched QCD.
The first one is a conventional analysis  in which  we
investigate the volume dependence of the energy shift for the ground state.
Different volume dependence is expected for scattering 
and bound states.
In the second one we investigate the energy level of the first excited state
employing the diagonalization method~\cite{Luscher:1990ck}
to separate the  first excited state from the ground state near the threshold of 2$m_N$.
If we find the ground state slightly below the threshold and the first excited state slightly above it,
then such a configuration of the two lowest levels is consistent with the ground state 
being a bound state and the first excited state a scattering state with almost zero relative 
momentum.
This method was previously used in a scalar QED simulation
to distinguish a system with or without a bound state~\cite{Sasaki:2006jn}.

Hereafter we call the analyses employed in the first and second calculations 
the single state and two state analyses, respectively.
The results in this article have been reported in Ref.~\cite{Yamazaki:2011xx}.

\section{Single state analysis}
\label{sec:1st}

We generate quenched configurations with the 
Iwasaki gauge action~\cite{Iwasaki:1983cj} at $\beta = 2.416$ whose
lattice spacing is $a=0.128$ fm, corresponding to $a^{-1} = 1.541$ GeV,
determined with $r_0=0.49$ fm 
as an input~\cite{AliKhan:2001tx}.
We take three lattice sizes, 
$L^3\times T = 24^3 \times 64$, $48^3 \times 48$ and $96^3 \times 48$, 
to investigate the spatial volume dependence of the energy 
difference between the two-nucleon ground state and twice  the nucleon mass.
The physical spatial extents are 3.1, 6.1 and 12.3 fm, respectively.

We use the tadpole improved Wilson action 
with $c_{\mathrm{SW}} = 1.378$~\cite{AliKhan:2001tx}.
Since it becomes harder to obtain a reasonable signal-to-noise ratio at
lighter quark masses for the multi-nucleon system, 
we employ a heavy quark mass at $\kappa = 0.13482$ which gives
$m_\pi = 0.8$ GeV for the pion mass and $m_N = 1.6$ GeV for the nucleon mass. 
Statistics is increased by repeating the measurement of 
the correlation functions
with the source points in different time slices on each configuration.

The quark propagators are solved with
the periodic boundary condition 
in all the spatial and temporal directions
using the exponentially smeared source
$
q^\prime(\vec{x},t) = \sum_{\vec{y}} A\, e^{-B|{\vec x} - \vec{y}|} q(\vec{y},t)
$
after the Coulomb gauge fixing.
On each volume we employ two sets of   
smearing parameters: $(A,B) = (0.5,0.5)$, $(0.5,0.1)$ 
for $L=24$ and $(0.5,0.5)$, $(1.0,0.4)$ for $L=48$ and 96.
The onset of ground state can be confirmed by 
consistency of effective masses with different sources as shown later. 
Hereafter the nucleon operators using the first and the second smearing 
parameter sets are referred to as ${\cal O}_{1}$ and ${\cal O}_{2}$, 
respectively.

\begin{figure}[!t]
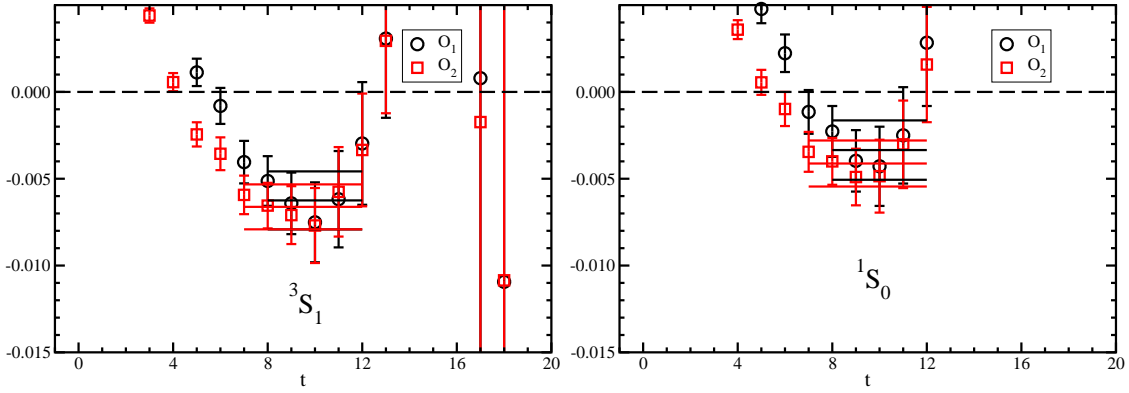

\includegraphics*[angle=0,width=0.49\textwidth]{Fig/eff_R_NN-3S1.eps}
\includegraphics*[angle=0,width=0.49\textwidth]{Fig/eff_R_NN-1S0.eps}
\caption{
Effective energy shifts for $^3$S$_1$ (left) and $^1$S$_0$ (right) channels.
Circle and square denote results for the ${\cal O}_{1,2}$ source operators, 
respectively.
\label{fig:eff_R}
}
\end{figure}
\subsection{Numerical results}

In order to determine the energy shift $\Delta E_L=E_{NN}-2M_N$ 
precisely in each volume, 
we define the ratio of the two-nucleon correlation function divided by
the nucleon correlation function squared,
$
R(t) = G_{NN}(t)/(G_N(t))^2,
$
where the same source operator is chosen for $G_{\mathrm{NN}}(t)$ 
and $G_N(t)$.
The effective energy shift is extracted as
$
\Delta E_L^{\mathrm{eff}} = \ln (R(t)/R(t+1)).
$

In the left panel of Fig.~\ref{fig:eff_R} we present typical results of 
time dependence of $\Delta E_L^{\mathrm{eff}}$ for 
the ${\cal O}_{1,2}$ sources in the $^3$S$_1$ channel, both of which
show negative values 
beyond the error bars in the plateau region of $t=8$--11.
Note that this plateau region is reasonably consistent 
with that for the effective mass
of the two-nucleon correlation functions.
The signals of $\Delta E_L^{\mathrm{eff}}$ 
are lost beyond $t\approx 12$ because of 
the large fluctuations in the two-nucleon correlation functions.
We determine $\Delta E_L$ by an exponential fit of the ratio in 
the plateau region, $t=8$--12 for ${\cal O}_1$ and 
$t=7$--12 for ${\cal O}_2$, respectively.
We obtain a similar quality for the signal for 
the $^1$S$_0$ channel on the (6.1 fm)$^3$ box
shown in the right panel of Fig.~\ref{fig:eff_R}.

\begin{figure}[!t]
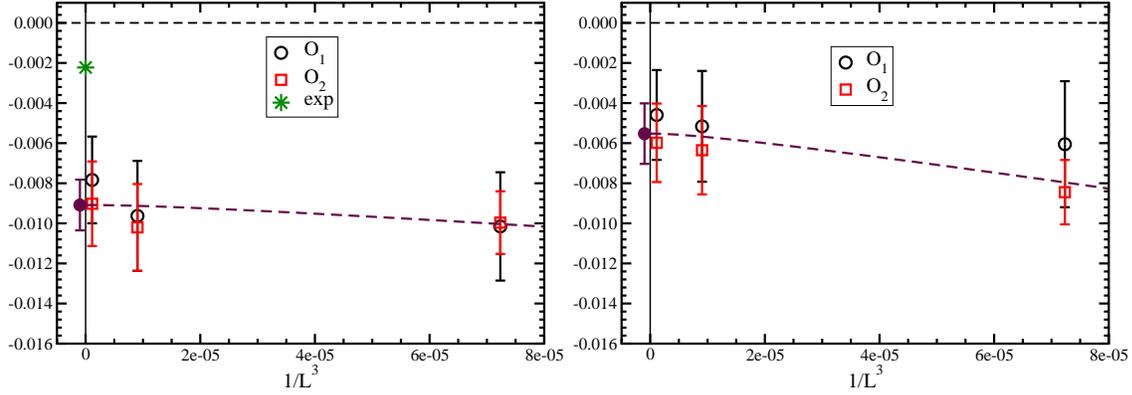

\includegraphics*[angle=0,width=0.49\textwidth]{Fig/dE_31.eps}
\includegraphics*[angle=0,width=0.49\textwidth]{Fig/dE_10.eps}
\caption{
Volume dependence of energy shifts $\Delta E_L$ as function of $1/L^3$ for
$^3$S$_1$ (left) and $^1$S$_0$ (right) channels.
The dashed line is extrapolation to the infinite volume,
and filled circle denotes the extrapolated result.
\label{fig:dE}
}
\end{figure}

The volume dependence of the energy shift $\Delta E_L$ 
for the $^3$S$_1$ channel
is plotted as a function of $1/L^3$ in the left panel of Fig.~\ref{fig:dE}. 
The results for the ${\cal O}_{1,2}$ sources are consistent 
within the error bars.
Little volume dependence for $\Delta E_L$ indicates 
a bound state, rather than the $1/L^3$ dependence expected for a 
scattering state, for the ground state in the $^3$S$_1$ channel.
The binding energy in the infinite spatial volume limit 
is extracted by a simultaneous fit of the data for the ${\cal O}_{1,2}$ sources 
employing 
the fit function including a finite volume effect for the 
two-particle bound state~\cite{Sasaki:2006jn,Beane:2003da},
\begin{equation}
\Delta E_L = -\frac{\gamma^2}{m_N}\left\{
1 + \frac{C_\gamma}{\gamma L} \sum^{\hspace{6mm}\prime}_{\vec{n}}
\frac{\exp(-\gamma L \sqrt{\vec{n}^2})}{\sqrt{\vec{n}^2}}
\right\},
\end{equation}
where $\gamma$ and $C_\gamma$ are free parameters, 
$\vec{n}$ is three-dimensional integer vector, 
and $\sum^\prime_{\vec{n}}$ denotes the summation without $|\vec{n}|=0$.
The binding energy, $-\Delta E_\infty$, is determined from 
$
-\Delta E_\infty = -\gamma^2/m_N,
$
assuming
$
2\sqrt{m_N^2 - \gamma^2} - 2 m_N \approx -\gamma^2/m_N.
$
The systematic error is estimated from the difference of the central values
of the fit results choosing different fit ranges
in the determination of $\Delta E_L$, and also using a constant fit as an alternative fit 
form.   Adding the statistical and systematic errors by quadrature, 
we obtain $-\Delta E_\infty =9.1(1.3)$ MeV for the binding energy. 
From the result, we conclude that the ground state in the $^3$S$_1$ channel
is a bound state.

The right panel of Fig.~\ref{fig:dE} plots the volume dependence 
of the energy shift $\Delta E_L$ 
for the $^1$S$_0$ channel.
Employing the same analysis as in the $^3$S$_1$ channel,
we find that  $-\Delta E_\infty = 5.5(1.5)$ MeV 
in the infinite volume limit,
which is 3.7 $\sigma$ away from zero.
This tells us that the ground state in the $^1$S$_0$ channel
is also bound at $m_\pi = 0.8$ MeV.
Since the existence of the bound state in this channel is not expected
at the physical quark mass, it might be 
a consequence of much heavier quark mass used in our calculation.

\section{Two-state analysis}
\label{sec:2nd}

The focus of the analysis is the characteristic feature, well known from quantum mechanics, 
that the existence of a bound state 
implies a negative scattering length, and hence 
a scattering state just above the two-particle threshold in a finite volume.  
Our investigation is carried out with the diagonalization~\cite{Luscher:1990ck} 
of 2$\times$2 correlation function matrix.

We work with two spatial extents, 4.1 fm and 6.1 fm. 
The corresponding lattice sizes are $L^3\times T = 32^3\times 48$
and $48^3 \times 48$, respectively.
The latter is the same size as in the first ensemble, but 
we regenerate independent configurations.
Most of the simulation parameters, including the gauge and fermion
actions, lattice spacing, quark mass, are identical to those
explained in the previous section.  
The diagonalization method for the 2$\times$2 matrix requires 
two operators each at source and sink time slice, which are
explained in our full paper~\cite{Yamazaki:2011xx}.
We note that due to our criteria of the choice of the operators,
this analysis does not provide an independent check for 
the ground state energy against the single state analysis.

We diagonalize the following matrix at each $t$,
$
M(t,t_0) = C(t_0)^{-1}C(t),
$
where $C(t)$ is the correlation function matrix of
the two-nucleon operators and
$t_0$ a reference time.
We determine the two eigenvalues $\lambda_\alpha(t)$ ($\alpha=0,1$) 
of $ M(t,t_0)$ at each $t$ and extract the energy of each eigenstate $\alpha$ 
through $\lambda_\alpha(t) =\exp (-\overline{E}_{L,\alpha}(t-t_0))$.
In order to determine the energy shift from the threshold
as in the single state analysis,
we define the ratio of the eigenvalue obtained from the diagonalization
to the nucleon correlation function squared,
$
\overline{R}_{\alpha}(t) = 
\lambda_{\alpha}(t)/(G_N(t))^2.
$
We also define the effective energy shift of the ratio 
$\overline{R}_{\alpha}$ as,
$\Delta \overline{E}_{L,\alpha}^{\mathrm{eff}} = 
\ln (\overline{R}_{\alpha}(t)/
{\overline{R}_{\alpha}(t+1)}).
$

\begin{figure}[!t]
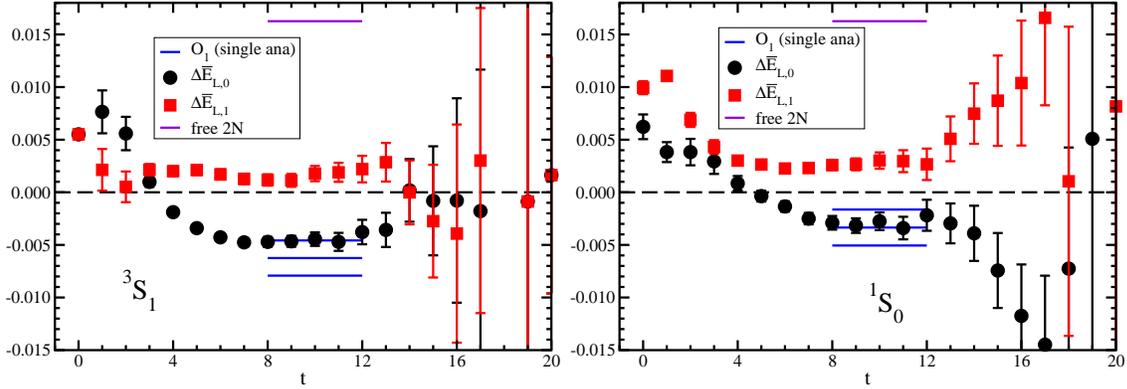

\includegraphics*[angle=0,width=0.49\textwidth]{Fig/eff_R_D3S1.eps}
\includegraphics*[angle=0,width=0.49\textwidth]{Fig/eff_R_D1S0.eps}
\caption{
Effective energy shifts $\Delta \overline{E}_{L,\alpha}^{\mathrm{eff}}$ 
obtained from eigenvalues for ground (circle) and 
first excited (square) states in $^3$S$_1$ (left) and $^1$S$_0$ (right) 
channels.
Three solid lines and single line denote the result of the single
state analysis with the ${\cal O}_1$ operator 
and free two-nucleon energy with the lowest momentum.
\label{fig:eff_R_D}
}
\end{figure}

\subsection{Numerical results}

We show the results for the $^3$S$_1$ channel on the (6.1 fm)$^3$ box.
The effective energy shifts for the $^3$S$_1$ channel 
are plotted in the left panel of Fig.~\ref{fig:eff_R_D}.
In Fig.~\ref{fig:eff_R_D} we use the nucleon correlator
with the ${\cal O}_1$ operator for the denominator of 
$\overline{R}_{\alpha}(t)$.
The ground state result $\Delta \overline{E}_{L,0}^{\mathrm{eff}}$
is reasonably consistent with the result of 
the single state analysis with the ${\cal O}_1$ source, 
which is expressed by the three solid lines in the figure.
The first excited state is clearly higher than the ground state, but it 
is much lower than the free case with the lowest relative momentum, whose
energy is given by $2\sqrt{m_N^2 + (2\pi/L)^2}$ 
denoted by the single solid line in the figure.

In the right panel of Fig.~\ref{fig:eff_R_D}
the effective energy shifts for the ground and first excited states
$\Delta \overline{E}_{L,\alpha}^{\mathrm{eff}}$ for the $^3$S$_1$ channel
are shown.
We find features similar to those in the $^3$S$_1$ channel.  
We observe that the absolute value of the energy shift of the ground state
is almost half of that in the 
$^3$S$_1$ channel.  This is consistent with the observation in the first
calculation.
On the other hand, the energy shift of the first excited state  
shown in the figure is
almost twice larger than that in the $^3$S$_1$ channel.
This finding is consistent with the property of a system which contains
a shallow bound state: The scattering length negatively increases as the binding energy decreases,  diverging when the binding energy vanishes.
From the results we confirm that the two-nucleon system in the both
channels at the heavy quark mass of $m_\pi = 0.8$ GeV
has a bound state.

\begin{figure}[!t]
\hfil
\includegraphics*[angle=0,width=0.49\textwidth]{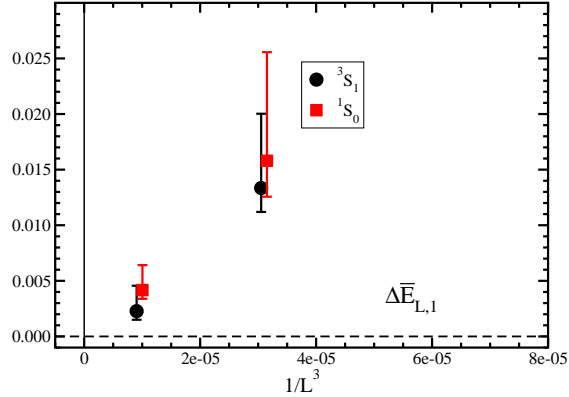}
\caption{
Volume dependence of energy shift $\Delta \overline{E}_{L,1}$
obtained from the eigenvalues
of the first excited state for $^3$S$_1$ and $^1$S$_0$ channels
as a function of $1/L^3$. The statistical and systematic errors are
added in quadrature.
\label{fig:DdE_n1}
}
\end{figure}

In Fig.~\ref{fig:DdE_n1} we plot the energy shift for the first excited state from the two lattice 
volumes as a function of $1/L^3$.  A roughly linear behavior, 
with a larger shift  on the (4.1 fm)$^3$ box compared to a smaller shift on the
 (6.1 fm)$^3$ box,  is consistent with this state being a scattering state. 
We evaluate the scattering length using L\"uscher's finite volume formula~\cite{Luscher:1986pf}, 
where we find reasonable consistency between the two volumes~\cite{Yamazaki:2011xx}.  
If our finding of a bound state in quenched QCD at heavy 
quark mass smoothly continues to the physical point, then this is the first 
calculation which explained a negative scattering length for the deuteron channel.

\section{Conclusion and discussion}
\label{sec:summary}

We have carried out two calculations in quenched QCD to investigate
whether the two nucleon systems
have a bound state or not at the heavier quark mass, corresponding
to $m_\pi = 0.8$ GeV.
In the first calculation, we have focused on the ground state
of the two-nucleon system, and have investigated the volume dependence 
of the energy shifts obtained with two different source operators.
In the second calculation we have carried out two-state analysis using
the diagonalization method with the $2\times 2$ correlation function matrix. 
Based on these results we have concluded that the ground state is
a bound state at the heavy quark mass in both the channels.

While similar bound state in the $^1$S$_0$ channel 
was observed in recent $N_f = 2+1$ QCD 
calculation at $m_\pi = 390$ MeV~\cite{Beane:2011xx},
the existence of the bound state looks odd from the experimental point of view.
We expect that the bound state vanishes at some lighter quark mass, 
where the scattering length diverges changing the sign 
from negative to positive. 
Further reduction of the quark mass would decrease the scattering length.
Confirmation of this scenario requires to investigate the quark mass
dependences of the binding energy and the scattering length.
We leave this study to future work.

\section*{Acknowledgments}
Numerical calculations for the present work have been carried out
on the HA8000 cluster system at Information Technology Center
of the University of Tokyo, on the PACS-CS computer 
under the ``Interdisciplinary Computational Science Program'' of 
Center for Computational Sciences, University of Tsukuba, 
and on the T2K-Tsukuba cluster system at University of Tsukuba. 
We thank our colleagues in the PACS-CS Collaboration for helpful
discussions and providing us the code used in this work.
This work is supported in part by Grants-in-Aid for Scientific Research
from the Ministry of Education, Culture, Sports, Science and Technology 
(Nos. 18104005, 18540250, 22244018) and 
Grants-in-Aid of the Japanese Ministry for Scientific Research on Innovative 
Areas (Nos. 20105002, 21105501, 23105708).

\bibliography{yamazaki}

\providecommand{\href}[2]{#2}\begingroup\raggedright\begin{thebibliography}{10}

\bibitem{Yamazaki:2009ua}
{\bf PACS-CS} Collaboration, T.~Yamazaki, Y.~Kuramashi, and A.~Ukawa,  {\em Phys. Rev.} {\bf D81} (2010)
  111504(R).
%, [\href{http://xxx.lanl.gov/abs/0912.1383}{{\tt arXiv:0912.1383}}].

\bibitem{Beane:2010hg}
{\bf NPLQCD} Collaboration, S.~R. Beane {\em et.~al.},  
  {\em Phys. Rev. Lett.} {\bf 106} (2011) 162001.

\bibitem{Beane:2011xx}
{\bf NPLQCD} Collaboration, S.~R. Beane {\em et.~al.},  
  arXiv:1109.2889[hep-lat].

\bibitem{Inoue:2010es}
{\bf HALQCD} Collaboration, T. Inoue {\em et.~al.},  
  {\em Phys. Rev. Lett.} {\bf 106} (2011) 162002.

\bibitem{Yamazaki:2011xx}
{\bf PACS-CS} Collaboration, T.~Yamazaki, Y.~Kuramashi, and A.~Ukawa,  {\em Phys. Rev.} {\bf D84} (2011)
  054506.
%, [\href{http://xxx.lanl.gov/abs/0912.1383}{{\tt arXiv:0912.1383}}].

\bibitem{Fukugita:1994na}
M.~Fukugita, Y.~Kuramashi, H.~Mino, M.~Okawa, and A.~Ukawa,
  {\em Phys. Rev. Lett.} {\bf 73} (1994) 2176;
%  [\href{http://xxx.lanl.gov/abs/hep-lat/9407012}{{\tt hep-lat/9407012}}];
M.~Fukugita, Y.~Kuramashi, M.~Okawa, H.~Mino, and A.~Ukawa,  
  {\em Phys. Rev.} {\bf D52} (1995).
%  3003--3023, [\href{http://xxx.lanl.gov/abs/hep-lat/9501024}{{\tt
%  hep-lat/9501024}}].

\bibitem{Beane:2006mx}
S.~R. Beane, P.~F. Bedaque, K.~Orginos, and M.~J. Savage,  
  {\em Phys. Rev. Lett.} {\bf 97} (2006) 012001.
%  [\href{http://xxx.lanl.gov/abs/hep-lat/0602010}{{\tt hep-lat/0602010}}].

\bibitem{Beane:2009py}
{\bf NPLQCD} Collaboration, S.~R. Beane {\em et.~al.},  
  {\em Phys. Rev.} {\bf D81} (2010) 054505.
%  [\href{http://xxx.lanl.gov/abs/0912.4243}{{\tt arXiv:0912.4243}}].

\bibitem{Ishii:2006ec}
N.~Ishii, S.~Aoki, and T.~Hatsuda,
  {\em Phys. Rev. Lett.} {\bf 99} (2007) 022001;
%  [\href{http://xxx.lanl.gov/abs/nucl-th/0611096}{{\tt nucl-th/0611096}}].
S.~Aoki, T.~Hatsuda, and N.~Ishii,
  {\em Prog. Theor. Phys.} {\bf 123} (2010) 89;
S.~Aoki, T.~Hatsuda, and N.~Ishii, 
  {\em Comput. Sci. Dis.}  {\bf 1} (2008) 015009
%\bibitem{Aoki:2008hh}
%S.~Aoki, T.~Hatsuda, and N.~Ishii, {\it {Nuclear Force from Monte Carlo
%  Simulations of Lattice Quantum Chromodynamics}},  {\em Comput. Sci. Dis.}
%  {\bf 1} (2008) 015009, [\href{http://xxx.lanl.gov/abs/0805.2462}{{\tt
%  arXiv:0805.2462}}].

\bibitem{Luscher:1986pf}
M.~L{\"u}scher,  {\em Commun. Math. Phys.}
  {\bf 105} (1986) 153;
M.~L{\"u}scher,  {\em Nucl. Phys.} {\bf B354} (1991) 531.

\bibitem{Luscher:1990ck}
M.~L{\"u}scher and W.~Wolff,  {\em Nucl. Phys.}
  {\bf B339} (1990) 222.

\bibitem{Sasaki:2006jn}
S.~Sasaki and T.~Yamazaki,  {\em Phys. Rev.} {\bf D74} (2006) 114507.
%  [\href{http://xxx.lanl.gov/abs/hep-lat/0610081}{{\tt hep-lat/0610081}}].

\bibitem{Beane:2003da}
S.~R. Beane, P.~F. Bedaque, A.~Parreno, and M.~J. Savage,  
  {\em Phys. Lett.} {\bf B585} (2004) 106--114.
%  [\href{http://xxx.lanl.gov/abs/hep-lat/0312004}{{\tt hep-lat/0312004}}].

%\bibitem{Beane:2007qr}
%S.~R. Beane, W.~Detmold, and M.~J. Savage, {\it {n-Boson Energies at Finite
%  Volume and Three-Boson Interactions}},  {\em Phys. Rev.} {\bf D76} (2007)
%  074507, [\href{http://xxx.lanl.gov/abs/0707.1670}{{\tt arXiv:0707.1670}}].

\bibitem{Iwasaki:1983cj}
Y.~Iwasaki, {Report No. UTHEP-118 (1983) (unpublished)}.
%\bibitem{Iwasaki:1983cj}
%Y.~Iwasaki and T.~Yoshie, {\it Renormalization group improved action for su(3)
%  lattice gauge theory and the string tension},  {\em Phys. Lett.} {\bf B143}
%  (1984) 449.

\bibitem{AliKhan:2001tx}
{\bf CP-PACS} Collaboration, A.~Ali~Khan {\em et.~al.},  {\em
  Phys. Rev.} {\bf D65} (2002) 054505.
%  [\href{http://xxx.lanl.gov/abs/hep-lat/0105015}{{\tt hep-lat/0105015}}].

%\bibitem{Omelyan:2003om}
%I.~P. Omelyan, I.~M. Mryglod, and R.~Folk, {\it {}},  {\em Comput. Phys.
%  Commun.} {\bf 151} (2003) 272--277.

%\bibitem{Takaishi:2005tz}
%T.~Takaishi and P.~de~Forcrand, {\it {Testing and tuning new symplectic
%  integrators for hybrid Monte Carlo algorithm in lattice QCD}},  {\em Phys.
%  Rev.} {\bf E73} (2006) 036706,
%  [\href{http://xxx.lanl.gov/abs/hep-lat/0505020}{{\tt hep-lat/0505020}}].

%\bibitem{Beam:1967zz}
%J.~E. Beam, {\it {Orthogonal Classification of Alpha-Particle Wave Functions}},
%   {\em Phys. Rev.} {\bf 158} (1967) 907--916.

%\bibitem{Bolsterli:1964zz}
%M.~Bolsterli and E.~Jezak, {\it {Vector Harmonics for Three Identical
%  Fermions}},  {\em Phys. Rev.} {\bf 135} (1964) B510--B515.

%\bibitem{Flambaum:2007mj}
%V.~V. Flambaum and R.~B. Wiringa, {\it {Dependence of nuclear binding on
%  hadronic mass variation}},  {\em Phys. Rev.} {\bf C76} (2007) 054002,
%  [\href{http://xxx.lanl.gov/abs/0709.0077}{{\tt arXiv:0709.0077}}].

\end{thebibliography}\endgroup

\end{document}